\documentclass[pre,groupedaddress,showkeys,showpacs,twocolumn]{revtex4}

\usepackage{amsmath}
\usepackage{graphics}
\usepackage{graphicx}
\usepackage{amsfonts}
\usepackage{amssymb}
\usepackage{dcolumn}
\usepackage{bm}
\usepackage{epstopdf}
\begin{document}
\title{Effects of stretching on the frictional stress of rubber.}
\author{Antoine Chateauminois, Danh-Toan Nguyen and Christian Fr\'etigny}
\email [] {antoine.chateauminois@espci.fr}
\affiliation{Soft Matter Sciences and Engineering Laboratory (SIMM),PSL Research
University, UPMC Univ Paris 06, Sorbonne Universités, ESPCI Paris, CNRS, 10 rue Vauquelin,
75231 Paris cedex 05, France}
\date{\today}
\begin{abstract}
In this paper, we report on new experimental results on the effects of in-plane surface stretching on the friction of Poly(DiMethylSiloxane) (PDMS) rubber with smooth rigid probes. Friction-induced displacement fields are measured at the surface of the PDMS substrate under steady-state sliding. Then, the corresponding contact pressure and frictional stress distributions are determined from an inversion procedure. Using this approach, we show that the local frictional stress $\tau$ is proportional to the local stretch ratio $\lambda$ at the rubber surface. Additional data using a triangular flat punch indicate that $\tau(\lambda)$ relationship is independent on the contact geometry. From friction experiments using pre-stretched PDMS substrate, it is also found that the stretch-dependence of the frictional stress is isotropic, i.e. it does not depend on the angle between stretching and sliding directions. Potential physical explanations for this phenomenon are provided within the framework of Schallamach's friction model. Although the present experiments are dealing with smooth contact interfaces, the reported $\tau(\lambda)$ dependence is also relevant to the friction of statistically rough contact interfaces, while not accounted for in related contact mechanics models.
\end{abstract}
\maketitle
%
\section{Introduction}
\label{sec:Introduction}
In many practical situations, soft solids such as elastomers, gels or biological tissues experience mechanical loading. These systems being very easily deformed, even the lightest stresses can induce strain level well beyond the small strain hypothesis, which remains an open issue for the description of their mechanical behaviour. Such situations are especially encountered in many contact experiments involving rigid probes where, as a consequence of the finite size of the contacting bodies, high in-plane strains are invariably experienced at the periphery of the contact.  As an example, one can cite the friction of rubber with spherical glass probes which was found to result in local strain higher than 30\%.~\cite{barquins1985,nguyen2011} Another example is the peeling of soft adhesive tapes where large tensile stresses are combined with localized friction processes within the regions located at the peeling front.~\cite{newby1997,newby1995}\\
\indent As far as one is concerned with the friction of rubber materials, one do not a priori expect friction to be affected by stretching by virtue of the liquid-like nature of these systems well above their glass transition temperature. However, some scarce experimental observations tend to suggest the opposite. In a couple of papers dealing with pre-stressed rubbers strips,~\cite{barquins1991,barquins1993} Barquins and co-workers concluded that tensile stretching affect rolling friction with rigid cylinders. Although controversial,~\cite{gay2000} their explanation was based on the hypothesis of a decrease in adhesion energy of the stretched rubber. They also showed that pre-stretching could affect sliding friction but in a regime complicated by the occurrence Schallamach detachment waves.~\cite{barquins1976} More recently, Yashima \textit{et al}~\cite{yashima2015} observed that frictional shear stress within smooth contact between silicone rubber and glass spherical probes could depend on contact size or on the curvature of the contact interface, an effect which could tentatively be related to difference in contact-induced surface strains.\\
\indent More generally, these overlooked issues pertain to any local physical description of rubber friction, especially with statistically rough surfaces. In such systems, individual micro-asperity contacts occur locally on a rubber surface which is known to experience finite stretch gradient at the scale of the macroscopic contact.~\cite{barquins1985,nguyen2011} However, the effects of such finite strains on micro-contacts shape and stresses are largely overlooked in current mean field descriptions of rough contacts (see e.g. \cite{scaraggi2015}), although they may affect the prediction of the actual contact area and the associated frictional forces.\\
\indent In the present study, we report on new experimental evidences of a dependence of frictional stress on surface stretch within macroscopic sliding contacts between a smooth silicone rubber and rigid probes. From optical contact imaging methods, we determine both surface displacement and stress fields with a space resolution of about 10~$\mu$m. Using this approach, we show that the local frictional stress is proportional to the local in-plane surface stretch independently on contact pressure, sliding velocity and contact geometry. Additional experiments using pre-stressed silicone strips also show that the increase in frictional stress with stretch ratio is isotropic, i.e. it is insensitive to the orientation of the sliding direction with respect to the stretch direction. Potential explanations for this phenomenon are discussed within the framework of the Schallamach's model of rubber friction.
%
\section{Experimental details}
\label{sec:experimental_details}
Friction experiments are carried out using either unstretched or uniaxially stretched Poly(DiMethylSiloxane) (PDMS) substrates in contact with smooth glass lenses. Contacts with the unstretched PDMS specimens are investigated using a custom-built device allowing to measure optically friction-induced surface displacement fields within the contact zone with sub-micrometer accuracy and a space resolution of about 10~$\mu$m. As detailed in reference,~\cite{nguyen2011} surface displacements are measured from the deformation of a square network of small cylindrical holes stamped on the PDMS surface by means of standard soft lithography techniques Then, a numerical inversion procedure allows to retrieve the corresponding contact stress distribution while taking into account the geometrical and material non linearities associated with the occurrence of finite strains within the contact (see Nguyen~\textit{et al}~\cite{nguyen2011} for further details). These experiments are carried out under a constant applied normal force (between 1.4 and 3.3~N) and at imposed sliding velocity (between 0.1 and 1~mm~s$^{-1}$) using BK7  plano-convex glass lenses with radii of curvature ranging from 5.2 to 25.9~mm.\\
\indent Additionally, some experiments with unstretched PDMS specimens are performed using a triangular flat punch which was micro-machined from a Poly(MethylMetAcrylate) (PMMA) block. The contacting face of the punch has the shape of an isoceles right triangle with two 6~mm edges. In order to minimize stress concentration at the edge of the contact, the corners of the triangular punch are rounded with a 300~$\mu$m radius of curvature.\\
\begin{figure}[ht]
	\begin{center}
		\includegraphics[width=0.8\columnwidth]{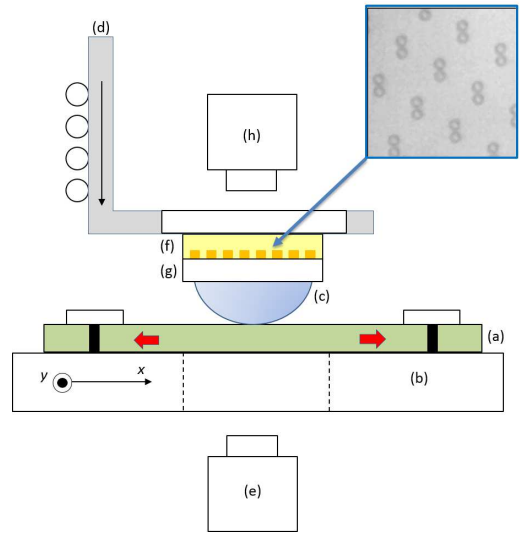}
		\caption{Schematic description of the custom-built experiments for friction measurements on uniaxially pre-stretched PDMS substrates. A stretched PDMS strip (a) is clamped on two crossed linear translation stages (b) allowing to vary the sliding direction with respect to the stretching axis. Contact with a plano-convex glass lens (c) is achieved under an imposed normal displacement condition using a manual linear translation stage (d).  A CMOS camera (e) allows contact visualization trough the transparent PDMS substrate. The frictional force along the sliding axis is measured using a custom made sensor consisting in a silicone disk (f) enclosed between a patterned (with SU8 resin pillars) glass disk (g) and a glass plate. After stiffness calibration, the force is determined from sub-pixel measurements of the displacement of the pattern using a CMOS camera (h). The inset shows two superimposed images of the pillars pattern taken before and after the application of a frictional force.} 
		\label{fig:setup}
	\end{center}
\end{figure}
\indent Using another dedicated set-up, friction experiments using uniaxially pre-stretched PDMS substrates are also carried out  in order to investigate whether a bulk tensile stretching could result in an anisotropy in the frictional shear stress. As schematically described in Fig.~\ref{fig:setup}, a PDMS strip is stretched between two grips on a loading frame which is fixed on two crossed motorized translation stages which allow to vary the angle between stretching and sliding directions. Contact is achieved under an imposed displacement condition using a glass lens with a radius of curvature of 5.2~mm. The friction force along the sliding direction is measured using a custom-built optical load sensor. It consists in a thin (1~mm) layer of a silicone elastomer crosslinked between a flat glass slide and a glass disk 20~mm in diameter on top of which the glass lens is glued. The internal face of the glass disk in contact with the silicone rubber is marked with a network of cylindrical posts (diameter 20~$\mu$m, center-to-center spacing 70~$\mu$m) using conventional SU8 resin lithography. It is easily detected optically as a result of the mismatch between the refractive index of the SU8 resin and the silicone material. The displacement of this markers' network under shear is measured by digital image correlation techniques and translated into a frictional force through appropriate calibration. The geometrical confinement of the incompressible silicone layer was intended to prevent tilting motions during shearing. In addition, the stiffness of the load sensor can be tuned by playing with the shear modulus of the silicone elastomer. Based on a work by Palchesko~\textit{et al},~\cite{palchesko2012} this can be achieved by mixing in various weight ratios commercially available Sylgard 184 and 527 PDMS (Dow Chemicals, Midland, MI). Here, mixing of Sylgard 184 and Sylgard 527 in a 40:60 weight ratio resulted in an isotropic lateral stiffness of 15.3~10$^{3}$~N~m$^{-1}$.\\
\indent Sylgard 184 PDMS is used as an elastomer substrate for all friction experiments. The silicon monomer and the hardener are mixed in a 10:1 weight ratio and crosslinked at $70\,^{\circ}\mathrm{C}$ for 48 hours. Contact experiments with unstretched PDMS are carried out using $15~\times~30~\times~60$~mm$^{3}$ specimens marked with a square network of small cylindrical holes (diameter 20~$\mu$m, depth 2~$\mu$m and center-to-center spacing 400~$\mu$m) in order to measure the surface displacement field under steady-state sliding conditions.\\
\indent Friction experiments with uniaxially stretched PDMS substrates are performed using $5~\times~30~\times~100$~mm$^{3}$ strips. The stretch ratio of the PDMS substrate in the contact region is measured optically from the deformation of a holes square network (diameter 20 $\mu$m, depth 5 $\mu$m and center-to-center spacing 80 $\mu m$) at the surface of the rubber specimen.\\
\indent In all the experiments to be reported, the contact conditions ensured the achievement of semi-infinite contact conditions (\textit{i.e.} the ratio of the substrate thickness to the contact radius is larger than ten \cite{gacoin2006}).\\
%
%
\section{Stretch and stresses within contact area}
\label{sec:stretch_stress}
\begin{figure}[ht] 
	\begin{center}
		\includegraphics[width=0.8\columnwidth]{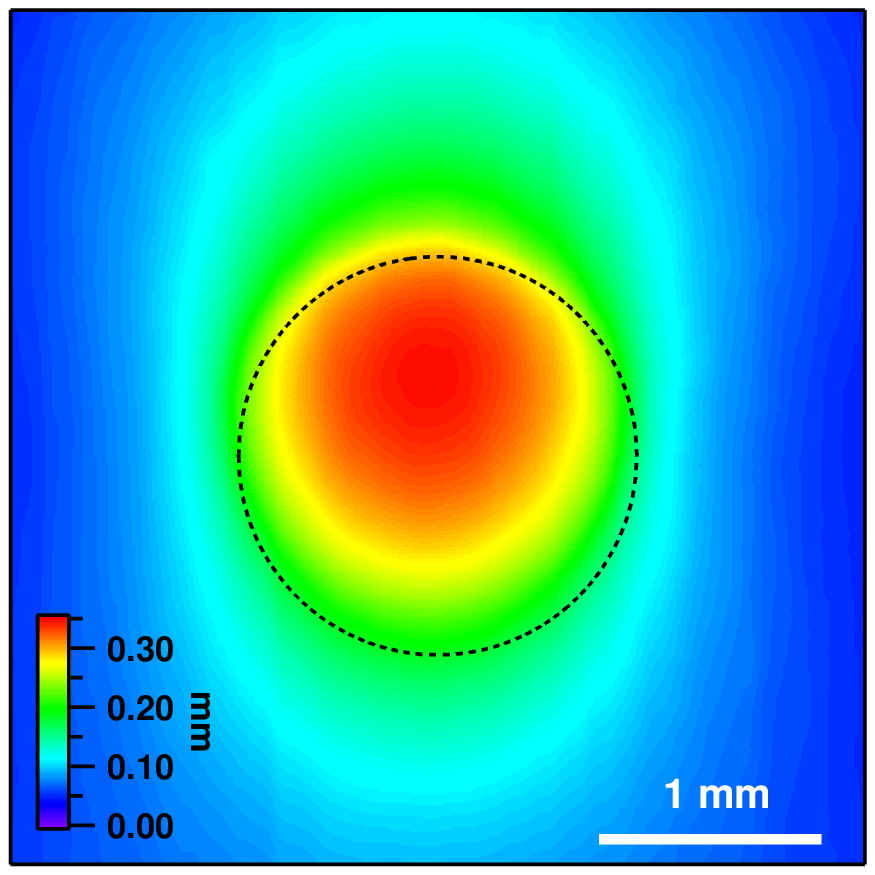}\\
		\includegraphics[width=0.8\columnwidth]{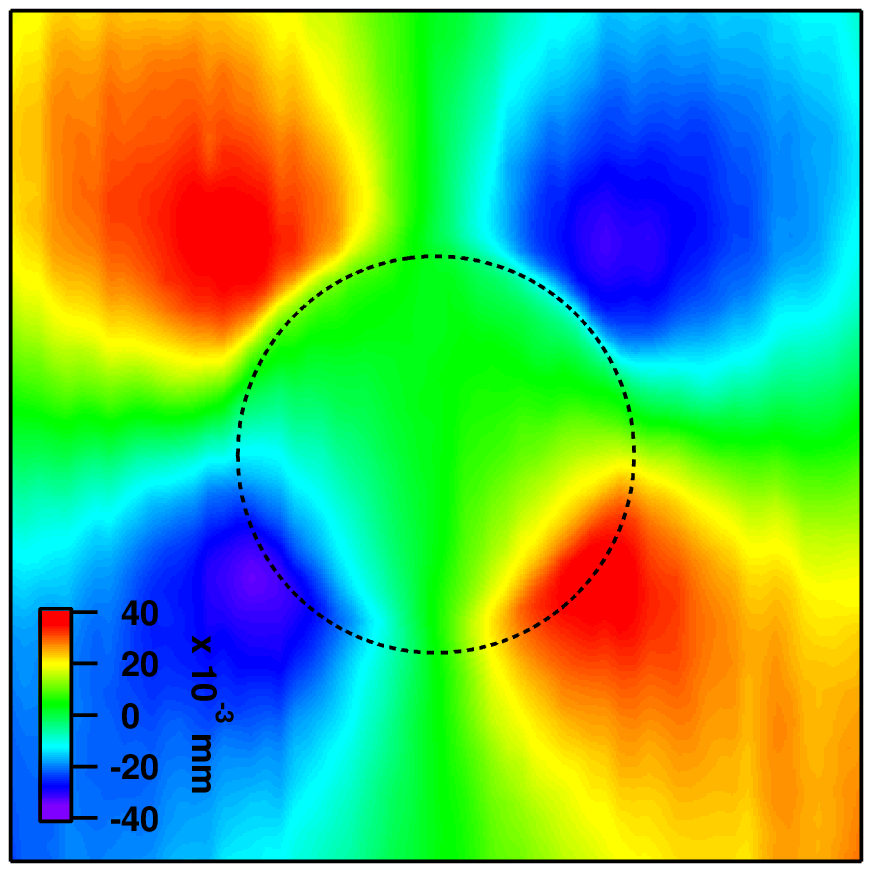}
		\caption{Measured displacement field during steady-state sliding of a glass lens (radius of curvature $R=9.3$~mm) on a PDMS substrate (sliding velocity $v=0.5$~mm~s$^{-1}$, applied normal load $P=1.4$~N). Top: displacement $u_y$ along the sliding direction; bottom: displacement $u_x$ perpendicular to the sliding direction. The PDMS substrate is moved from bottom to top with respect to the fixed glass lens. Displacements are mapped in Lagragian coordinates, i.e. relative to the equilibrium state before the application of lateral displacement. The dotted lines delimits the edge of the contact area using the same Lagragian coordinates. Note that the magnitude of $u_x$ is much lower than that of $u_x$.} 
		\label{fig:displ_field}
	\end{center}
\end{figure}

\begin{figure}[ht]
	\begin{center}
		\includegraphics[width=0.8\columnwidth]{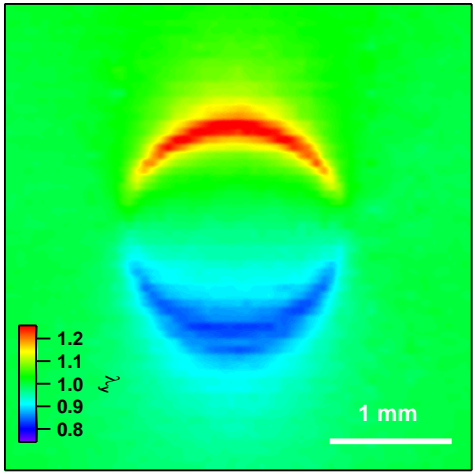}\\
		\includegraphics[width=0.8\columnwidth]{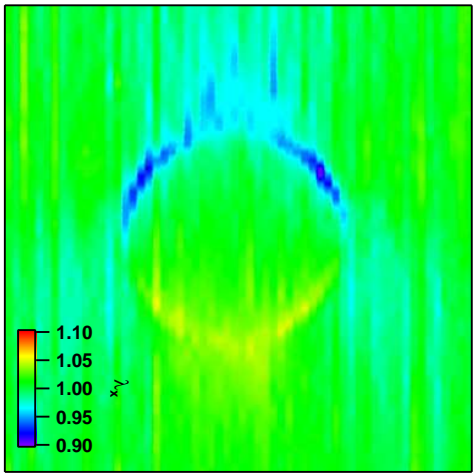}
		\caption{Measured stretch ratios along ($\lambda_y$, top) and perpendicular ($\lambda_x$, bottom) to the sliding direction in Lagragian coordinates (same experimental conditions as for Fig.~\ref{fig:displ_field}). The PDMS substrate is moved from bottom to top with respect to the fixed glass lens.} 
		\label{fig:lambda_field}
	\end{center}
\end{figure}

\begin{figure}[ht]
	\begin{center}
		\includegraphics[width=0.9\columnwidth]{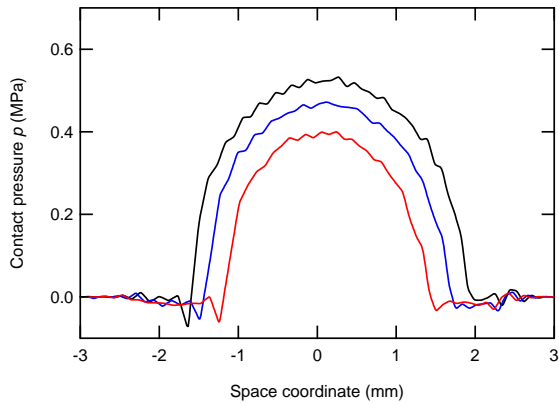}		
		\caption{contact pressure profiles across the contact zone and perpendicular to the sliding direction for increasing values of the applied normal load $P$ ($v=0.5$~mm~s$^{-1}$, $R=9.3$~mm). Red: $P=1.4$~N, blue: $P=2.3$~N, black: $P=3.3$~N.} 
		\label{fig:pressure_profile}
	\end{center}
\end{figure}

\begin{figure}[ht]
	\begin{center}
		\includegraphics[width=0.9\columnwidth]{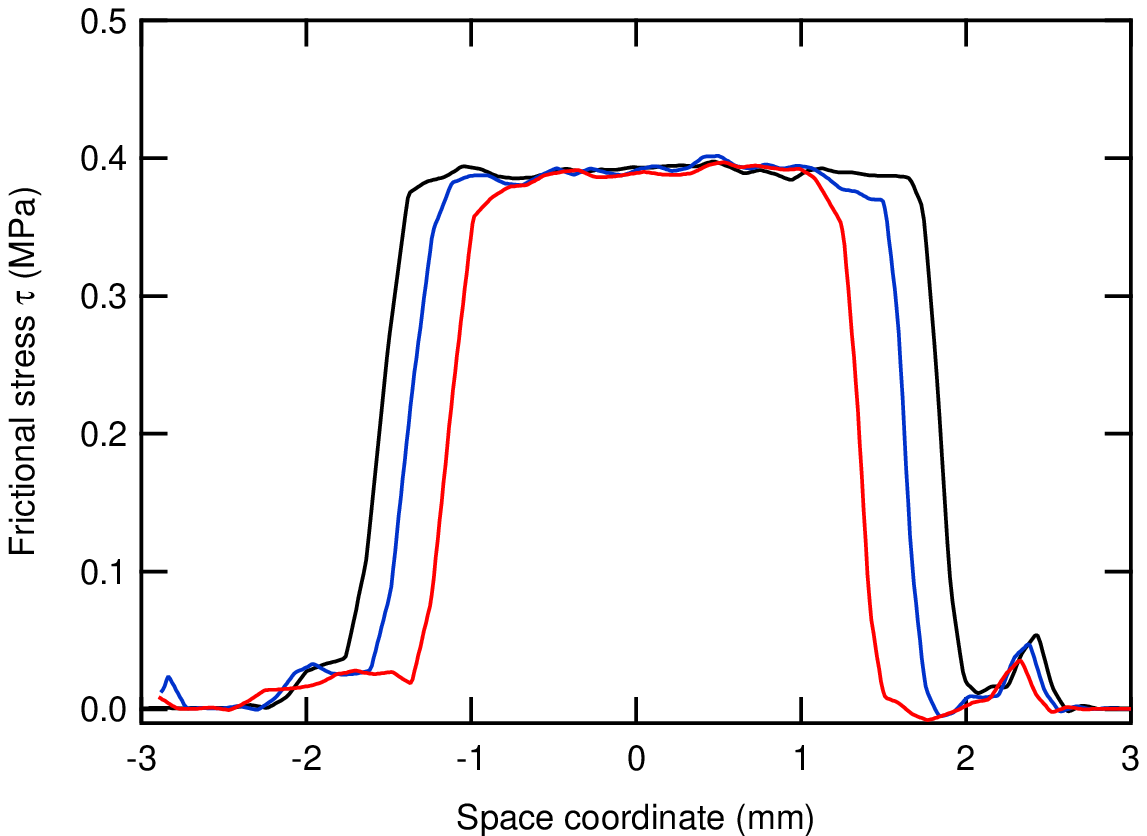}
		\includegraphics[width=0.9\columnwidth]{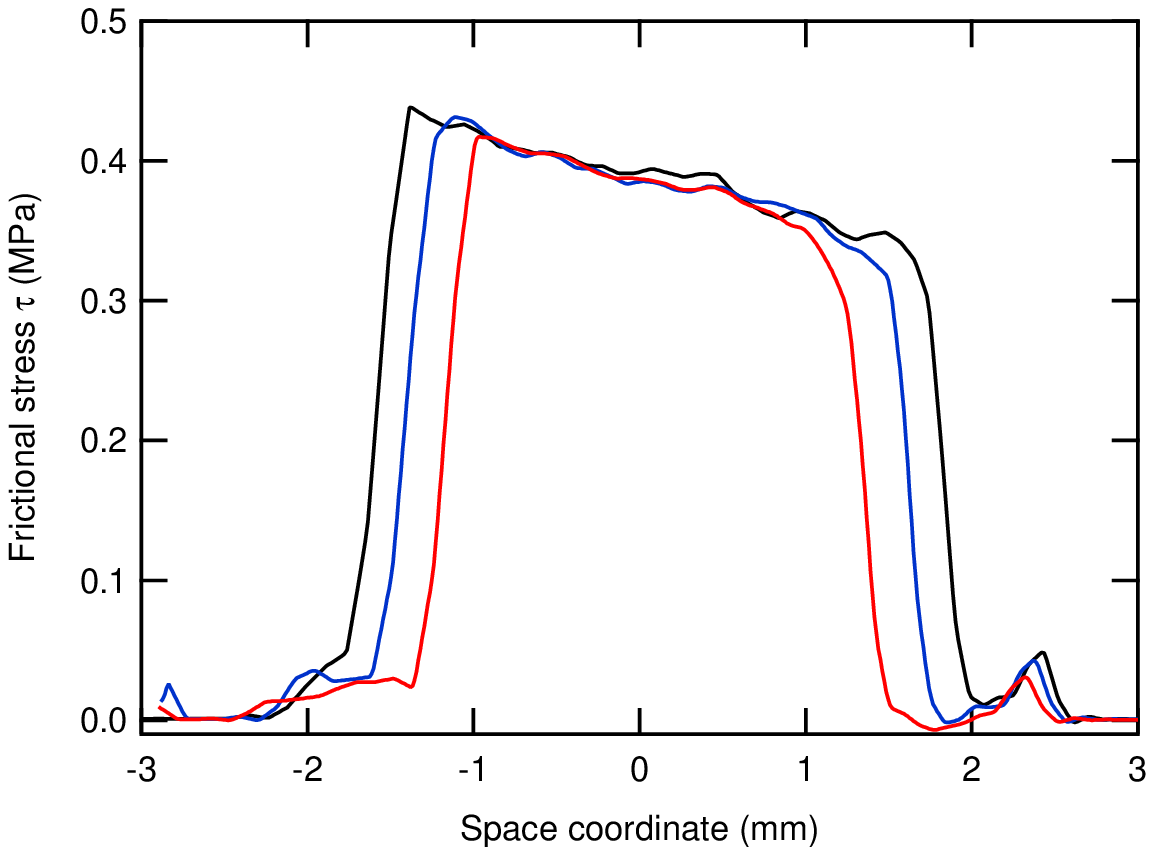}
		\caption{Shear stress profiles across the contact zone, both perpendicular to (top) and along (bottom) the sliding direction for increasing values of the applied normal load $P$ (same experimental conditions as for Fig.~\ref{fig:pressure_profile}). Red: $P=1.4$~N, blue: $P=2.3$~N, black: $P=3.3$~N.} 
		\label{fig:shear_profiles}
	\end{center}
\end{figure}

\begin{figure}[ht]
	\begin{center}
		\includegraphics[width=0.9\columnwidth]{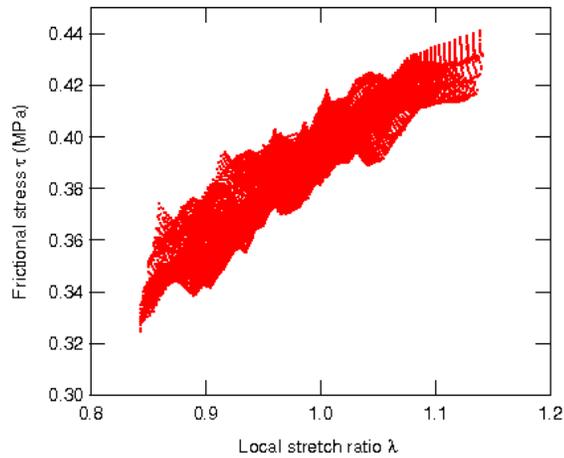}
		\caption{Local shear stress versus local stretch ratio within a sliding contact ($P=3.3$~N, $v=0.5$~mm~s$^{-1}$, $R=9.3$~mm).} 
		\label{fig:shear_lambda_field}
	\end{center}
\end{figure}

\begin{figure}[ht]
	\begin{center}
		\includegraphics[width=0.9\columnwidth]{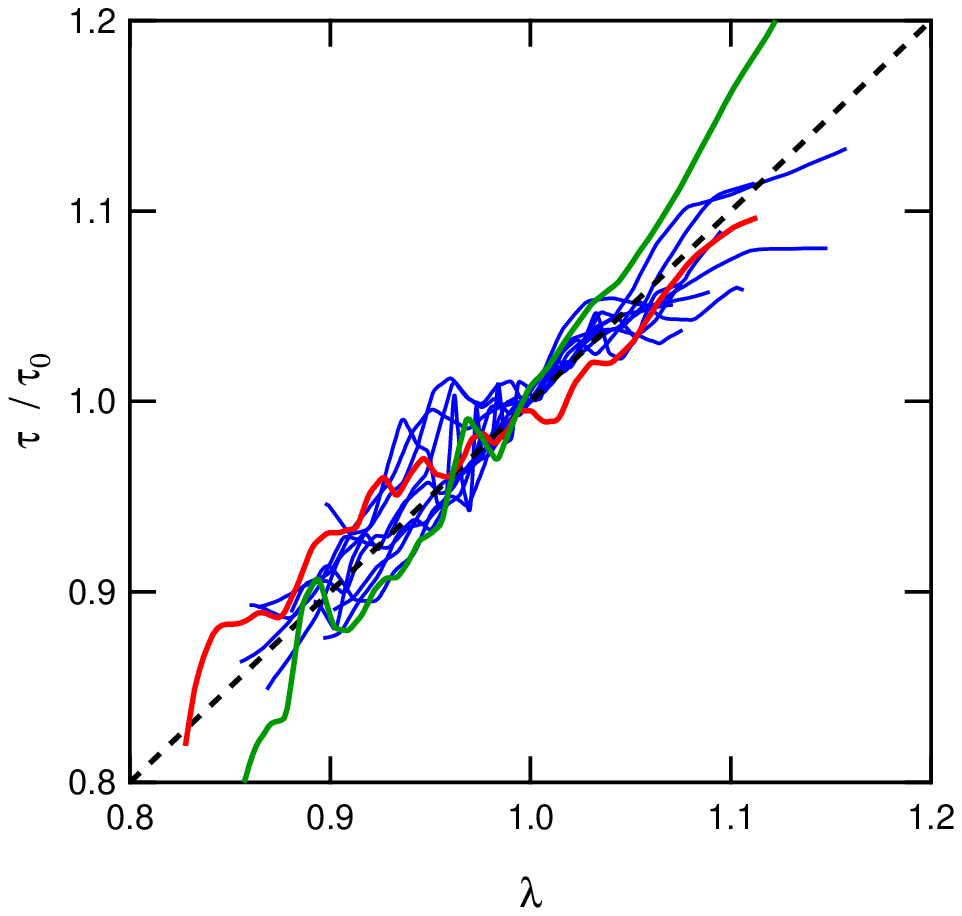}
		\caption{Normalized local frictional stress $\tau/\tau_0$ versus local stretch ratio $\lambda$ ($\tau_0$ is the measured local frictional stress for $\lambda=1$). Blue lines: spherical probes with $1.4<P<3.3$~N, $5.2<R<25.9$~mm and $0.1<v<1$~mm~s$^{-1}$. Red line: isoceles triangular flat punch with the sliding direction along the larger edge ($P=1.6$~N, $v=0.1$~mm~s$^{-1}$); green line: same with the sliding direction along the height of the isoceles triangle. For the shake of clarity, data were taken along contact cross-sections parallel to the sliding direction.} 
		\label{fig:tau_lambda}
	\end{center}
\end{figure}

\begin{figure}[ht]
	\begin{center}
		\includegraphics[width=0.9\columnwidth]{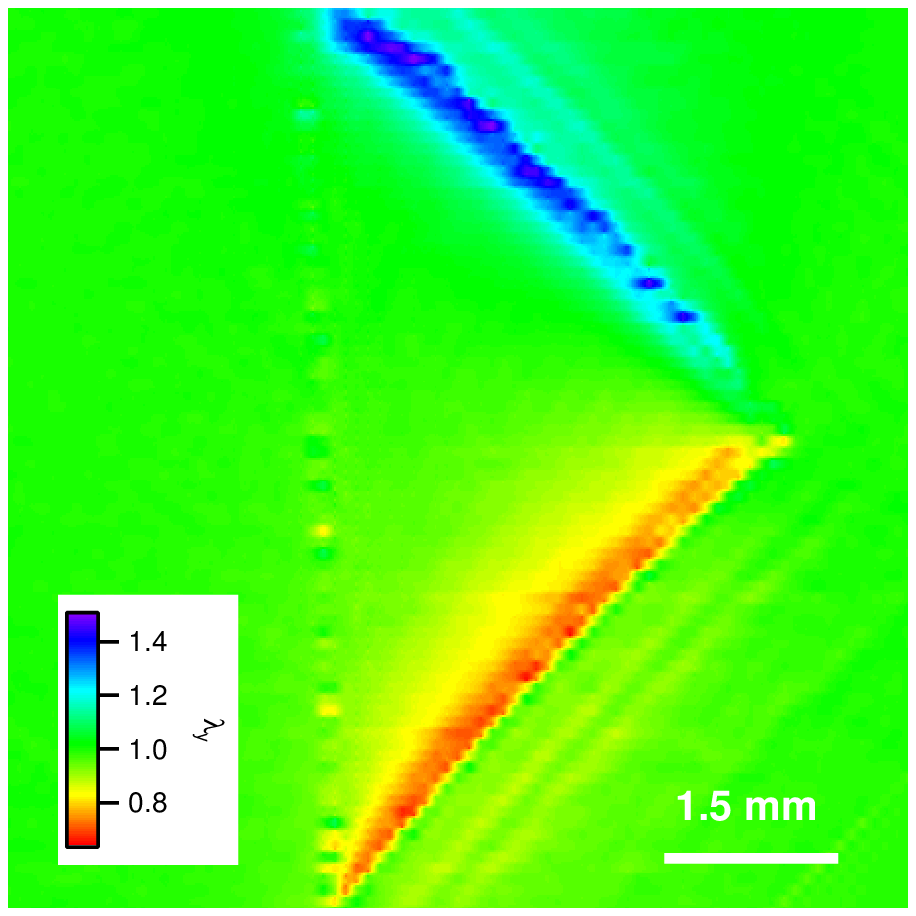}
		\includegraphics[width=0.9\columnwidth]{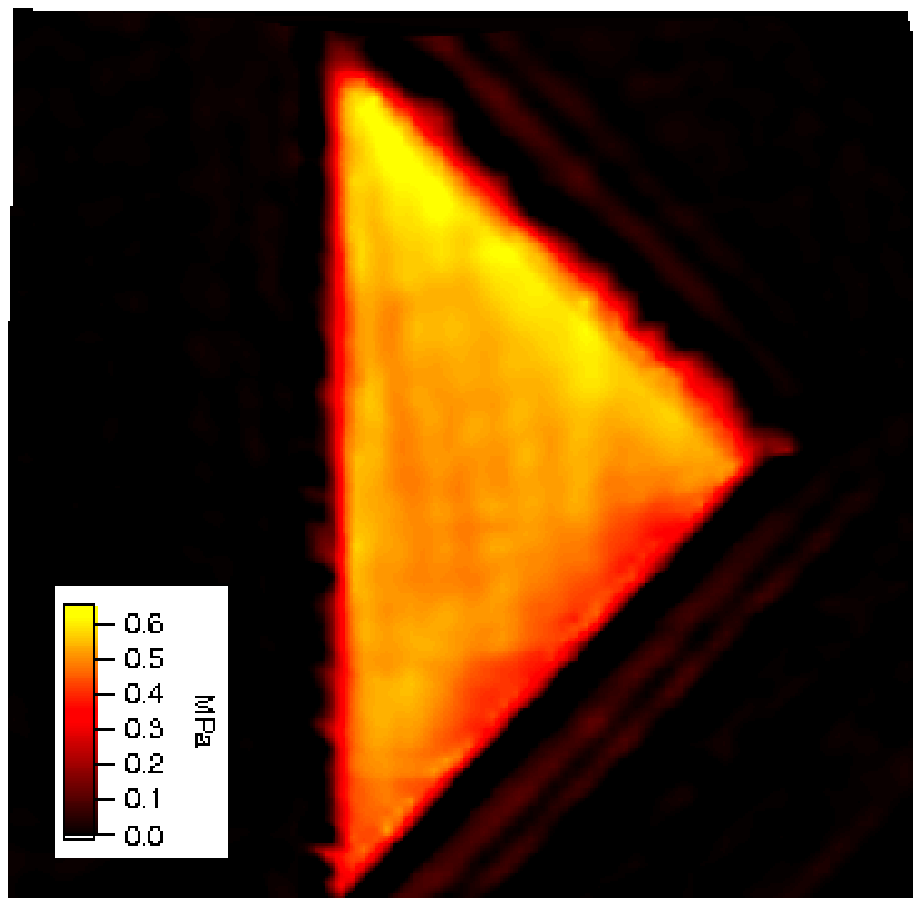}
		\caption{Longitudinal stretch ratio (top) and frictional shear stress (bottom) fields within a sliding contact between a PDMS substrate and a triangular flat punch ($v=0.1$~mm~s$^{-1}$, $P=1.6$~N). The PDMS substrate is moved from bottom to top with respect to the fixed flat punch.} 
		\label{fig:triangular_field}
	\end{center}
\end{figure}

Fig.~\ref{fig:displ_field} shows a typical example of a measured surface displacement field during steady-state friction of a spherical glass probe on a PDMS substrate. In this representation, in-plane displacements components $u_x$ and $u_y$ (where $y$ is the sliding axis and $x$ is perpendicular) are mapped in Lagragian coordinates, i.e. space coordinates refer to the deformation state prior to the application of lateral contact loading. The edge of the contact area is delimited in the same Lagrangian coordinates by a dotted line in this figure. The main displacement component is obviously along the sliding direction, the transverse displacements being about ten times lower. The quadrupolar symmetry of $u_x$ field reflects the occurrence of Poisson's effect: the PDMS surface is compressed (resp. stretched) along the sliding direction at the leading (resp. trailing) edge of the contact, which results in an expansion (resp. contraction) in the transverse direction.\\
\indent local surface stretching can be estimated from a space derivative of this displacement field. Fig.~\ref{fig:lambda_field}, shows the distribution of the longitudinal and transverse stretch ratio defined as $\lambda_y=1+\frac{\partial u_y}{\partial y}$ and $\lambda_x=1+\frac{\partial u_x}{\partial x}$, respectively. Transverse surface stretch $\lambda_x$ appears to be restricted to a very narrow band at the contact periphery and its magnitude is much lower than that of the longitudinal stretch $\lambda_y$. The later exhibits a clear gradient along the sliding direction, from compression at the leading edge of the contact to traction at the trailing edge. 
Noticeably, the maximum and minimum values of the longitudinal stretch ratio (about 1.27 and 0.83, respectively for the contact conditions under consideration in the figure) fall well beyond the small strain hypothesis. More precisely, previously reported tensile testing data for the used silicone rubber~\cite{fretigny2017} indicate that surface strains experienced by the substrate during steady state friction fall within the neo-Hokean range. Unless otherwise specified, all the stress distributions determined herein from the inversion of displacement fields were thus obtained using a neo-Hookean description of the mechanical response of PDMS.\\
\begin{figure}[ht]
	\begin{center}
		\includegraphics[width=0.8\columnwidth]{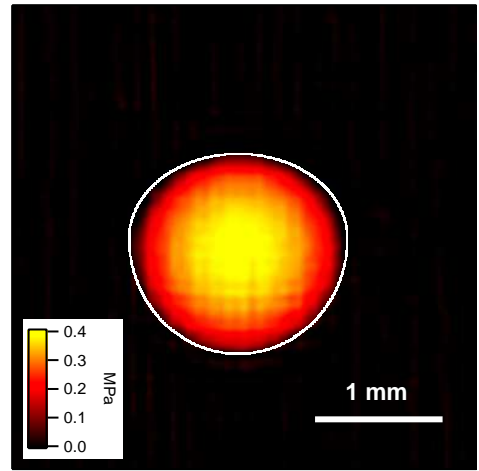}\\
		\includegraphics[width=0.8\columnwidth]{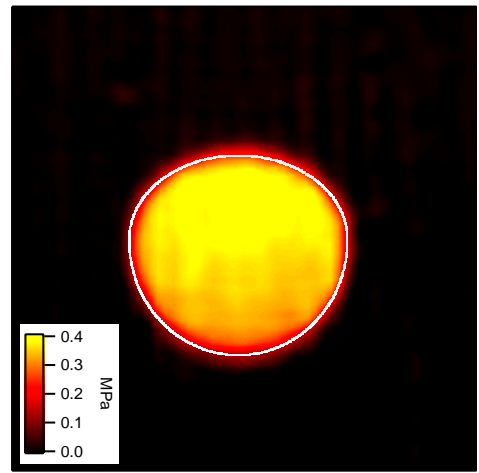}
		\caption{Contact pressure and frictional shear stress distribution deduced from the inversion of a measured displacement field (same experimental conditions as for Fig.~\ref{fig:displ_field}). The PDMS substrate is moved from bottom to top with respect to the fixed glass lens. Space coordinates are relative to the deformed state, i.e. stresses correspond to Cauchy's stresses expressed in Eulerian coordinates. The white lines delimits the edge of the contact area in the same deformed space.} 
		\label{fig:stress_field}
	\end{center}
\end{figure}
\indent Fig.~\ref{fig:stress_field} details the characteristic contact pressure and frictional shear stress distributions which are obtained from the inversion of measured surface displacement fields under steady-state friction with a smooth spherical probes. Here, according to an Eulerian description, we refer to Cauchy stress components, i.e. to equilibrium stresses expressed in the deformed space. As detailed in Fig.~\ref{fig:pressure_profile}, contact stress distribution exhibits a classical Hertzian shape which do not deserve further comments. In what follows, we focus instead on the frictional stress field distribution which, at first sight, seems constant for the smooth contact interface under consideration.\\
\indent However, a careful examination of the corresponding stress field reveals a clear gradient of the frictional stress along the sliding direction with a minimum at the leading edge of the contact and a maximum at the trailing edge (Fig.~\ref{fig:shear_profiles}, bottom). Such a feature was systematically preserved whatever the contact pressure, the sliding velocity and the radius of curvature of the glass lens. A comparison of stress cross-sections perpendicular to (Fig.~\ref{fig:shear_profiles}, top) and along (Fig.~\ref{fig:shear_profiles}, bottom) the sliding direction indicates that the observed changes in frictional stress are not correlated to the contact pressure distribution. They can neither be explained from a simple viscoelastic effect which would induce a stress gradient in the opposite direction: shear stress would be higher in the less relaxed state encountered at the leading contact edge than in the more relaxed regions at the trailing edge. In addition, one can also mention that the frequency of the glass transition of the selected PDMS substrate at room temperature is about 10$^8$~Hz,~\cite{nguyen2013} while the characteristic strain frequency of the contact, $v/a$ (where $v$ is the sliding velocity and $a$ the contact radius) is no more than 10~Hz.\\
\indent Interestingly, it turns out that the frictional stress gradient is rather correlated to the value of the local longitudinal surface stretch ratio achieved within the contact. An example of this correlation is shown in Fig.~\ref{fig:shear_lambda_field} where the local frictional stress $\tau$ has been reported as a function of the local stretch ratio $\lambda=\lambda_y$ for a set of data points taken within the contact area ($\lambda$ values were transferred to Eulerian coordinates for that purpose). From this example, it turns out that the local frictional shear stress is linearly increasing from about 0.34 to 0.43~MPa when the local stretch ratio is increasing from 0.85 to 1.15.\\ 
\indent As shown in Fig.~\ref{fig:tau_lambda}, this linear increase in the frictional stress with stretch ratio is preserved whatever the radius of the spherical probe, the applied contact force and the sliding velocity within the ranges under consideration. For the shake of clarity, only $\tau(\lambda)$ data taken along contact cross-sections parallel to the sliding direction were reported as blue lines in this figure. In order to account for slight changes in the average frictional stress between different PDMS substrates and for different sliding velocities, the local frictional stress data $\tau$ where normalized with respect to the value $\tau_{0}$ corresponding to $\lambda=1$, i.e. to the unstretched state achieved at the middle of the contact area. From this plot, it emerges that the local frictional stress is always proportional to the local stretch ratio, i.e. $\tau=\lambda \tau_0$.\\
\indent Noticeably, the $\tau(\lambda)$ relationship is found to be independent on contact geometry. This is demonstrated by measurements using a flat triangular punch instead of a spherical probe. The corresponding stretch ratio and frictional shear stress fields are shown in Fig.~\ref{fig:triangular_field} with the sliding direction oriented along the largest edge of the triangular punch. Here again, a stretch gradient is evidenced along the sliding direction. Stress amplification at the vicinity of the edge of the flat punch is found to result in stretch ratios which are significantly higher (typically 1.45 a the trailing edge of the contact) than for spherical probes. As a result, inversion using a neo-Hookean behaviour law for the PDMS proved to be inaccurate and the numerical inversion of the displacement field was carried out using Yeoh's hyperelastic model~\cite{yeoh1993} which is able to account for the strengthening of the rubber response at high stretch ratios ($\lambda>1.3$).\\
\indent As seen in Fig.~\ref{fig:triangular_field}, the frictional stress is clearly increasing when the stretch ratio is changing from compression to traction at the leading and trailing contact edges, respectively. When reported in a $\tau/\tau_0(\lambda)$ representation (red line in Fig.~\ref{fig:tau_lambda}), it turns out that the stretch dependence of the frictional stress is exactly the same as for spherical probes. Moreover, this is verified whatever the orientation of the facets of the punch with respect to the sliding direction (the red and green lines in Fig.~\ref{fig:tau_lambda} correspond to sliding directions parallel and perpendicular, respectively, to the large edge of the triangular punch). As a conclusion, it emerges that the local frictional stress of the PDMS substrate sliding against a smooth rigid surface is proportional to the local value of the stretch ratio, independently of the geometry of the contact area, of the sliding velocity and of the applied contact pressure.
\section{Friction on a pre-stretched substrate}
\label{sec:pre_stretch}
\begin{figure}
	\begin{center}
		\includegraphics[width=1\columnwidth]{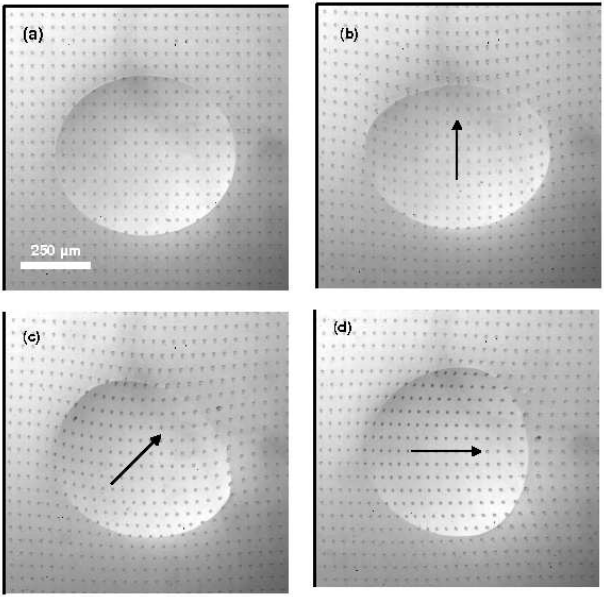}
		\caption{Images of the contact area between a pre-stretched ($\lambda_p=1.24$) PDMS substrate and a spherical probe ($R=5.2$~mm). Normal contact (a) and steady-state sliding ($v=0.1$~mm~s$^{-1}$)  with (b) $\theta=0$, (c) $\theta=-\pi/4$ and (d) $\theta=-\pi/2$ where $\theta$ is the angle between the stretching and sliding directions. The stretching direction is along the vertical direction and the displacement of the PDMS substrate with respect to the fixed glass lens is indicated by an arrow.} 
		\label{fig:stetched contacts}
	\end{center}
\end{figure}

\begin{figure}
	\begin{center}
		\includegraphics[width=1\columnwidth]{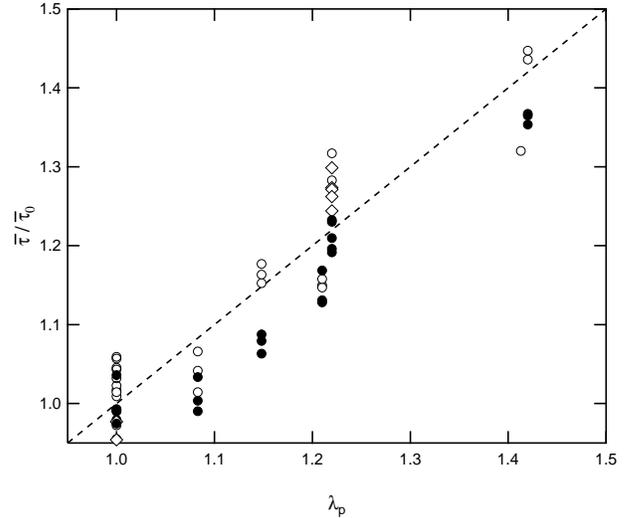}
		\caption{Normalized average frictional stress $\overline{\tau}/\overline{\tau_0}$ as a function of the tensile stretch ratio $\lambda_p$ of the PDMS substrate ($\overline{\tau}$ and $\overline{\tau_0}$ are the average frictional stress measured for $\lambda_p \ne 1$ and $\lambda_p=1$, respectively).($\circ$) $\theta=0$, ($\bullet$) $\theta=\pi/2$, ($\diamond$) $\theta=\pi/4$. } 
		\label{fig:tau_lambda_stretch}
	\end{center}
\end{figure}

Some additional insights into this dependence are provided by experiments where the stretch state of the rubber network is varied by superimposing a bulk tensile stretch $\lambda_p$ to the contact induced deformation. Such experiments especially allow to investigate whether the stretch dependence of the frictional stress depends on the orientation of the sliding direction with respect to the stretch direction. As fully detailed and accounted for in a companion paper~\cite{fretigny2017}, normal contact between a spherical probe and a stretched rubber substrate is characterized by an elliptical shape of the contact with the major axis perpendicular to the stretching direction. As an example, a normal contact image on a pre-stretched ($\lambda_p=1.24$) PDMS is show in Fig.~\ref{fig:stetched contacts}(a) where the ratio of the major to the minor axis of the elliptical contact is 1.12.\\
\indent Under steady-state sliding, strong changes in this initially elliptical shape of the contact area are observed which depend on the angle $\theta$ between the sliding and stretching directions (Fig.~\ref{fig:stetched contacts}). However, the associated frictional stress is observed to be independent on the sliding direction.
Here, we just consider the average value of the frictional stress $\overline{\tau}$ defined as the ratio of the frictional force $F$ to the actual contact area $A$ measured under steady state sliding. When the normalized value $\overline{\tau}/\overline{\tau_0}$ of this averaged shear stress (where $\overline{\tau_0}$ is the average shear stress of the unstretched rubber substrate) is reported as un function of the pre-stretch ratio $\lambda_p$ (Fig.~\ref{fig:tau_lambda_stretch}), we obtain exactly the same linear relationship as for the local $\tau(\lambda)$ analysis carried out using un-stretched substrates. Moreover, this dependence is found to be insensitive to the orientation of the sliding direction with respect to the stretch direction, despite a strong anisotropy in the contact shape.\\
%
%
\section{Discussion}
\label{sec:discussion}
From a local analysis of rubber friction with smooth rigid bodies, we showed that the local frictional shear stress is proportional to in-plane stretching of the elastomer surface. To our best knowledge, such a dependence has never been reported and its physical origin remains unclear. In order to get additional insights into this phenomenon, experiments were also carried out on uniaxially stretched substrates which showed that the stretch-dependence of friction is independent on sliding direction despite a strong anisotropy in the contact shape under steady-state sliding. It therefore turns out that this phenomenon is a generic feature of rubber friction.\\
\indent For the contact conditions under consideration, a weak, logarithmic, dependence of the frictional stress on velocity has previously been reported.\cite{nguyen2013} As a consequence, one cannot explain the observed changes in the local frictional stress from a consideration of the variations in local sliding velocity which result from the non-homogeneous deformation of the PDMS substrate: as a result of compressive (resp. tensile) surface strains at the leading (resp. trailing) edges of the contact, the actual sliding velocity at the contact interface is effectively lower (resp. higher) than the driving velocity imposed to the PDMS substrate. Here, the resulting variations in sliding velocity within the contact area (about $\pm$10\%) are too weak to account for the observed changes in the local frictional stress. In addition, such an explanation would not account for the effects of pre-stretching on the average frictional stress.\\
\indent In an attempt to explain the experimental results, one could go back to the molecular origins of rubber friction. Referring to the classical Schallamach model,~\cite{schallamach1963} the latter are assumed to be driven by thermally activated pinning/depinning events of elastomer molecules to the contacting surface. As detailed by Schallamach and others,\cite{Chernyack1986,leonov1990,schallamach1963,singh2011,vorvolakos2003} the rate of rupture of these bonds is likely to be enhanced on application of external stress and hence the process can be modelled using the Eyring rate equation. Such a description was found to successfully account for the experimental velocity dependence of rubber friction which is characterized by the existence of a peak.\cite{vorvolakos2003} As mentioned above, we are dealing experimentally with a sliding regime characterized by a logarithmic increase in the frictional stress with velocity. This regime would thus correspond to the low velocity side of the friction peak predicted by Schallamach's model. Previous calculations by Singh and Juvekar~\cite{singh2011} indicates that, in this regime, non-linearities in chain extension and finite chain extensibility does not affect significantly the frictional stress predicted by Schallamach's model. A similar conclusion also holds for viscous retardation of the chains.\\
According to Singh and Juvekar re-formulation of Schallamach model~\cite{singh2011}, a logarithmic velocity-dependence of the frictional stress $\sigma_0$ can be calculated in the form
\begin{equation}
\sigma_0 \approx \frac{N_0 k_BT}{2 \zeta} ln \frac{V_0}{uV^*}
\end{equation}
where $V_0$ is the steady velocity, $N_0$ represents the total number of surface sites to which the polymer can bond and $\zeta$ is an activation length, $u=e^{-W/k_BT}$ with $W$ the free energy difference between the bounded and unbounded states of the bonding site. $V*=k_BT/\tau \zeta M$ where $M$ represents the stiffness of the polymer chain and $\tau=\tau_0 e^{E/kT}$ with $E$ the bond activation energy.
This regime can be shown to exist if $u<<1$ and $u<V_0/V^*<1$.\\ 
According to the above expression, our experimental results can be explained within the framework of the Schallamach's model only if we make the assumption that the density of surface site available for bonding $N_0$ is proportional to the stretch ratio. However, for a liquid-like material such a rubber, one do not expect the chain density to vary with the extension of the surface. One should therefore postulate that surface stretching brings to the rubber surface some polar bonding sites which would otherwise remain buried within the bulk network and unavailable for pinning to the glass surface. Some analogies between such a mechanism and chemical modifications of PDMS surface may interestingly be drawn. As detailed in refs.,~\cite{lee2003,ng2002} exposing the PDMS surface to an air or oxygen plasma introduces silanol (Si-OH) groups and makes the surface hydrophilic. If the treated surface is left in contact with air, surface rearrangements progressively occur that bring new hydrophobic groups to the surface to lower the surface free energy and increase the wetting angle with water.\\
\indent Alternately, the observed stretch-dependence of friction could be considered within a continuum mechanics framework. If friction is still assumed to result from concentrated point forces acting on the rubber surface at some length scale, one could consider the effect of finite substrate strains on such forces. This can be accounted for by considering the formulation of surface Green's function for a stretched substrate which was recently derived by He~\cite{He2008} for neo-Hookean materials. By taking a concentrated force oriented to the axis 1, The Green's coefficient $G_{11}(\mathbf{x})$ provides the displacement of surface points situated along the same axis. For a stretch ratio $\lambda_1$ oriented along $x_1$, $G_{11}$ reads
\begin{equation}
G_{11}(x_1,0, 0)=\frac{\lambda_1}{2 \pi \mu x_1}
\end{equation}
This expression is similar to that derived for linear elasticity, expect that the shear modulus $\mu$ is replaced by an apparent modulus equal to $\mu/\lambda_1$. This result can be rationalized by considering that the above expression was derived using incremental deformation theory with a neo-Hookean consitutive law characterized by a decrease in the tangent modulus as a function of $\lambda_1$. However, it cannot not simply account for the experimental observation unless we speculate on the occurrence of some sliding heterogeneities, such as localized stick-slip events, at some intermediate length scale below the experimental space and time resolution. Finite strain effects such as those described by the above neo-Hookean Green's tensor could then be involved. In its spirit, such an approach would be reminiscent of spring-block models used to describe the propagation of slip fronts during sliding.~\cite{amundsen2015} Its validation would clearly require further experimental and theoretical investigations which are beyond the scope of this work.\\
\indent Although we provided some tentative explanations for the observed stretch-dependence of friction, it still need to be rationalized within the framework of a theoretical model. However, it must be emphasised that any theoretical description of the experimental results should account for the isotropy of the effect of substrate stretching on the frictional response 
\section{Conclusion}
In this study, we have shown that local frictional stress within smooth contacts between silicone rubber and rigid probes is proportional to the in-plane surface stretch of the elastomer substrate. Additionally, the observed increase in friction with stretch ratio is found to be isotropic, i.e. it does not depend on the relative orientation of sliding and stretching. To our best knowledge, such an effect has never been reported. Although the determination of its physical origin requires further elucidation, it is found to be very robust and independent on contact geometry. This overlooked stretch-dependence of rubber friction pertains to many different situations. As mentioned in the introduction, frictional contact with statistically rough surfaces are clearly an example where frictional stresses achieved locally at micro-contact scale will be affected by surface strains induced at the macroscopic contact scale. Interestingly, such surface strains are a consequence of the finite size of the contact. It therefore turns out that the stretch-dependence of friction introduces a coupling between the macroscopic and microscopic contact length scales which is clearly not accounted for in existing rough contact mechanics theories dealing with extended - infinite - contact interfaces.\\
The observed stretch-dependence of rubber friction also pertains to the contact response of bio-mimetic, fibrillar adhesives where friction occurs locally on slender elastomer fibrils which can be easily elongated well beyond linear regime.  
\section{Acknowledgments}
We gratefully acknowledge Chung Yuen Hui, Anand Yagota, Johan Le Chenadec and St\'ephane Roux  for very stimulating discussions about this paper. We are especially indebted to A. Jagota for suggesting us a potential effect of stretching on the availability of polar sites on the PDMS surface. The authors also wish to thank Alexis Prevost for his kind help in the realization of the triangular punch.\\
\bibliographystyle{unsrt}

\end{document}